\pdfoutput=1
\documentclass[5p]{elsarticle}
\usepackage[T1]{fontenc}
\usepackage[colorlinks=true,linkcolor=blue,urlcolor=blue,citecolor=blue]{hyperref}
\usepackage[]{lineno}
\usepackage{arydshln}
\usepackage[normalem]{ulem}

\begin{document}
\begin{frontmatter}
	
	\title{Excited electronic states of Sr$_2$: \textit{ab initio} predictions and experimental observation of the $2^1\Sigma^{+}_{u}$ state}

	\author[PAN]{Jacek Szczepkowski}\corref{mycorrespondingauthor}
	\ead{jszczep@ifpan.edu.pl}
	\cortext[mycorrespondingauthor]{Corresponding authors} 
	\author[FUWT]{Marcin Gronowski}
	\ead{Marcin.Gronowski@fuw.edu.pl}
	\author[PAN]{Anna Grochola}
	\author[PAN]{W{\l}odzimierz Jastrzebski}
	\author[FUWT]{Micha{\l} Tomza}
	\author[FUW]{Pawe{\l} Kowalczyk}
	
	\address[PAN]{Institute of Physics, Polish Academy of Sciences,
		al.~Lotnik\'{o}w~32/46, 02-668~Warsaw, Poland}
	\address[FUWT]{Institute of Theoretical Physics, Faculty of Physics, University of Warsaw, ul.~Pasteura~5, 02-093 Warszawa, Poland}
	\address[FUW]{Institute of Experimental Physics, Faculty of Physics,		University of Warsaw, ul.~Pasteura~5, 02-093~Warszawa, Poland}
	\begin{abstract}
		Despite its apparently simple nature with four valence electrons, the strontium dimer constitutes a challenge for modern electronic structure theory. Here we focus on excited electronic states of Sr$_2$, which we investigate theoretically up to 25000 cm$^{-1}$ above the ground state, to guide and explain new spectroscopic measurements. In particular, we focus on potential energy curves for the $1^1\Sigma^{+}_{u}$, $2^1\Sigma^{+}_{u}$, $1^1\Pi_{u}$, $2^1\Pi_{u}$, and $1^1\Delta_{u}$ states computed using several variants of advanced \textit{ab initio} methods to benchmark them. In addition, a new experimental study of the excited $2^1\Sigma^{+}_{u}$ state using polarisation labelling spectroscopy is presented, which extends knowledge of this state to high vibrational levels, where perturbation by higher electronic states is observed. The available experimental observations are compared with the theoretical predictions and help to assess the accuracy and limitations of employed theoretical models. The present results pave the way for future more accurate theoretical and experimental spectroscopic studies. 
	\end{abstract}
\end{frontmatter}

\section{Introduction}
\label{sec:introduction}

Diatomic molecules at ultralow temperatures are a perfect platform for research touching upon the very fundamentals of quantum physics and chemistry~\cite{CarrNJP09}. Ultracold polar molecules have been proposed and employed for a plethora of ground-breaking experiments ranging from quantum-controlled collisions and chemical reactions~\cite{BohnScience17} to quantum simulations~\cite{GrossScience17} and precision measurements of fundamental constants and their spatiotemporal variation~\cite{DeMilleScience17}. After spectacular successes with alkali-metal molecules, which can be efficiently formed from ultracold atoms using magnetoassociation~\cite{JulienneRMP06b} followed by optical stabilization~\cite{JulienneRMP06a}, the production of ultracold molecules containing alkaline-earth-metal atoms has emerged as another important research goal. 

Recently, an ultracold gas of Sr$_2$ dimers in their absolute ground state was obtained using all-optical methods, where weakly bound singlet-state molecules were formed in an optical lattice by narrow-line photoassociation and transferred to the ground rovibrational level by  the stimulated Raman adiabatic passage (STIRAP)~\cite{LeungNJP21}. Fast chemical reactions between such dimers were observed close to the universal limit. Nevertheless,  ultracold Sr$_2$ molecules have already been employed in a series of exciting experiments ranging from studying asymptotic physics in subradiant states~\cite{McGuyerNP15} to photodissociation with quantum state control~\cite{McdonaldNature16}. Very recently, a new type of molecular lattice clock based on ultracold Sr$_2$ dimers with long vibrational coherence has also been established~\cite{KondovNP19,Leung2022}. This paves the way for upcoming applications of these molecules in  quantum simulation~\cite{BhongalePRL13}, quantum metrology~\cite{SafronovaRMP18}, and precision measurements probing the fundamental laws of nature~\cite{ZelevinskyPRL08,KotochigovaPRA2009}.

Exciting developments and applications of ultracold molecules described above would not have been feasible without thorough experimental spectroscopic analysis and substantial theoretical \textit{ab initio} electronic structure evaluations of the underlying molecular structure. The required level of accuracy varies for each application. Generally, precise measurements can provide more accurate outcomes than theoretical calculations. However, \textit{ab initio} quantum-chemical calculations of potential energy curves, permanent and transition electric dipole moments, and other molecular couplings are frequently necessary to propose, guide, and explain experimental endeavors. 

Alkaline-earth-metal diatomic molecules, despite their apparently simple nature with four valence electrons and closed-shell ground electronic state, have constituted a challenge for modern electronic structure theory. Already the simplest Be$_2$ dimer presents unusually strong bonding and unique shape of the ground-state potential energy curve~\cite{MerrittScience09}, which accurate theoretical description required highly correlated methods~\cite{PatkowskiScience09}. Confirming the existence of elusive vibrational states of the ground-state Mg$_2$ dimer also needed state-of-the-art quantum-chemical calculations~\cite{YuwonoSA20}. Thus, it is not surprising that the accurate theoretical description of the Sr$_2$ dimer in the ground and excited electronic states may require careful treatment, similar to lighter neutral dimers or charged Sr$_2^+$ molecular ion~\cite{SmialkowskiJPB20}. 

The ground $X^1\Sigma^+_g$ and excited $2^1\Sigma^+_u$  and $ 3^1\Pi_u$  electronic states of Sr$_2$ were initially investigated experimentally with absorption and laser‐induced fluorescence spectroscopy~\cite{BergemanJCP80,Gerber1984,Bordas1992}, followed by high-resolution  Fourier-transform laser‐induced fluorescence spectroscopy of the $X^1\Sigma^+_g$ state~\cite{Stein2008,SteinEPJD10} and the minimum region of the excited  $1^1\Sigma^+_u$, $1^1\Pi_u$, and $2^1\Sigma^+_u$ states~\cite{Stein2011}. Recently, highly-accurate measurements with ultracold Sr$_2$ allowed for improving the accuracy of rovibrational spectra of the $X^1\Sigma^+_g$ and $1^1\Sigma^+_u$ states~\cite{LeungNJP21}. The ground and excited electronic states of Sr$_2$ were also investigated theoretically using different computational approaches, including large-core semiempirical pseudopotentials~\cite{Boutassetta1996,Czuchaj2003}, small-core relativistic pseudopotentials~\cite{SkomorowskiJCP12}, and all-electron relativistic Hamiltonian~\cite{Kotochigova2008}. The challenging character of calculations for excited molecular electronic states could be seen in contradictory dissociation energies for the lowest-excited $1^1\Sigma^+_u$ and $1^1\Pi_u$ states reported without detailed estimates of computational uncertainties. Several other studies focused solely on the ground $X^1\Sigma^+_g$ electronic state~\cite{WangJPCA00,MitinRJPCA09,YinJCP10,LiJPCA11,HeavenCPL11,YangTCA12}, which is already well understood.

In this work, we investigate the excited electronic states of the Sr$_2$ molecule. We start with the computational evaluation of the complete molecular electronic spectrum up to the excitation energy of around 25000$\,$cm$^{-1}$. Next, we compute potential energy curves for the $1^1\Sigma^{+}_{u}$, $2^1\Sigma^{+}_{u}$, $1^1\Pi_{u}$, $2^1\Pi_{u}$, and $1^1\Delta_{u}$ states using several variants of advanced \textit{ab initio} methods. New experimental measurements of the excited $2^1\Sigma^{+}_{u}$ state using polarisation labelling spectroscopy are presented, extending the range of observed vibrational levels to higher energies. The corresponding Dunham coefficients and experimental potential energy curve are reported. The observed perturbations in  the recorded spectrum give preliminary information on higher-excited electronic states. The comparison of the experimental observations with several theoretical predictions helps to assess and benchmark the accuracy and limitations of employed theoretical models.

The structure of the paper is the following. In Section~\ref{sec:theory}, we describe the employed theoretical methods and obtained computational results. In Section~\ref{sec:experiment}, we present the used experimental technique and the spectroscopic results. In section~\ref{sec:conclusion}, we conclude and provide an outlook.

\section{Electronic structure calculations }
\label{sec:theory}
\subsection{Computational methods}

Several computational approaches were used in the electronic structure calculations to assess and benchmark their accuracy. The all-electron computations employed the eXact-2-Component Hamiltonian~\cite{Peng2012} and the equation-of-motion coupled cluster method~\cite{Stanton1993} with single and double excitations (EOM-CCSD)~\cite{HAMPEL1992} with the relativistic correlation-consistent core-valence quadruple-zeta basis sets (aug-cc-pwCVQ-X2C)~\cite{bas2} in version implemented in the Molpro 2022.1 Program~\cite{molpro}. We explored the impact of the core-electron correlation on the results by correlating only valence electrons (denoted as x2cCCSDv), valence and $4s4p$ electrons (denoted as x2cCCSDbc), as well as valence, $4s4p$, and $3s3p3d$ electrons (denoted as x2cCCSDsc). 

Other equation-of-motion coupled cluster computations used the small-core relativistic energy-consistent ECP28MDF pseudopotential~\cite{Lim2006,DolgCR12} with the quadruple- and quintuple-zeta pseudopotential-based correlation-consistent polarised core-valence basis sets (aug-cc-pCVQZ-PP and aug-cc-pCV5Z-PP, donated as QZ and 5Z, respectively)~\cite{bas2} in the Cfour 2.1 software~\cite{cfour}.  We obtained the complete basis set (CBS) limit with two-point $1/X^3$ extrapolation~\cite{Helgaker1997}. To estimate the role of the higher excitations, we compared EOM-CCSD (denoted as ecpCCSD) and EOM-CCSDT-3~\cite{NOGA1987,He2001} (denoted as ecpCCSDT3). 

We also performed multireference computations with the standard Davidson correction~\cite{Davidson}. We described the valence-electron correlation by the multiconfiguration reference internally contracted configuration interaction method~\cite{iMRCI1988,iMRCI1988b,iMRCI1992} with the active space composed of 20 (sMRCI+Q) or 24 (MRCI+Q) orbitals. Such a sizeable active space is necessary to correctly describe the $2^1\Pi_u$ state of Sr$_2$. The orbitals were optimised at the complete active space self-consistent field (CASSCF) level~\cite{newMCSCF}. Additionally, we employed the hybrid CIPT2+Q method~\cite{CIPT2}, which adds the core-electron correlation to MRCI+Q by the multireference Rayleigh-Schrödinger second-order perturbation theory. All multireference computations used the aug-cc-pwCV5Z-PP basis set~\cite{bas2} and were performed with the Molpro 2022.1 program. Since we had a problem converging calculations for monomers in the dimer basis set, we assumed that the basis set superposition error for multireference methods is the same as for ecpCCSD. This assumption is well justified as the total basis set superposition error is relatively small for Sr$_2$.

We shifted the computed interaction energies by the sum of the appropriate experimentally-measured atomic excitation energies from the NIST database~\cite{NIST} and 1081.64~cm$^{-1}$, corresponding to the molecular ground state's depth~\cite{SteinEPJD10}. This procedure guarantees that the reported energies are relative to the ground state's minimum and tend to the corresponding atomic values in their asymptotes.

\begin{figure*}
\includegraphics[width=\textwidth]{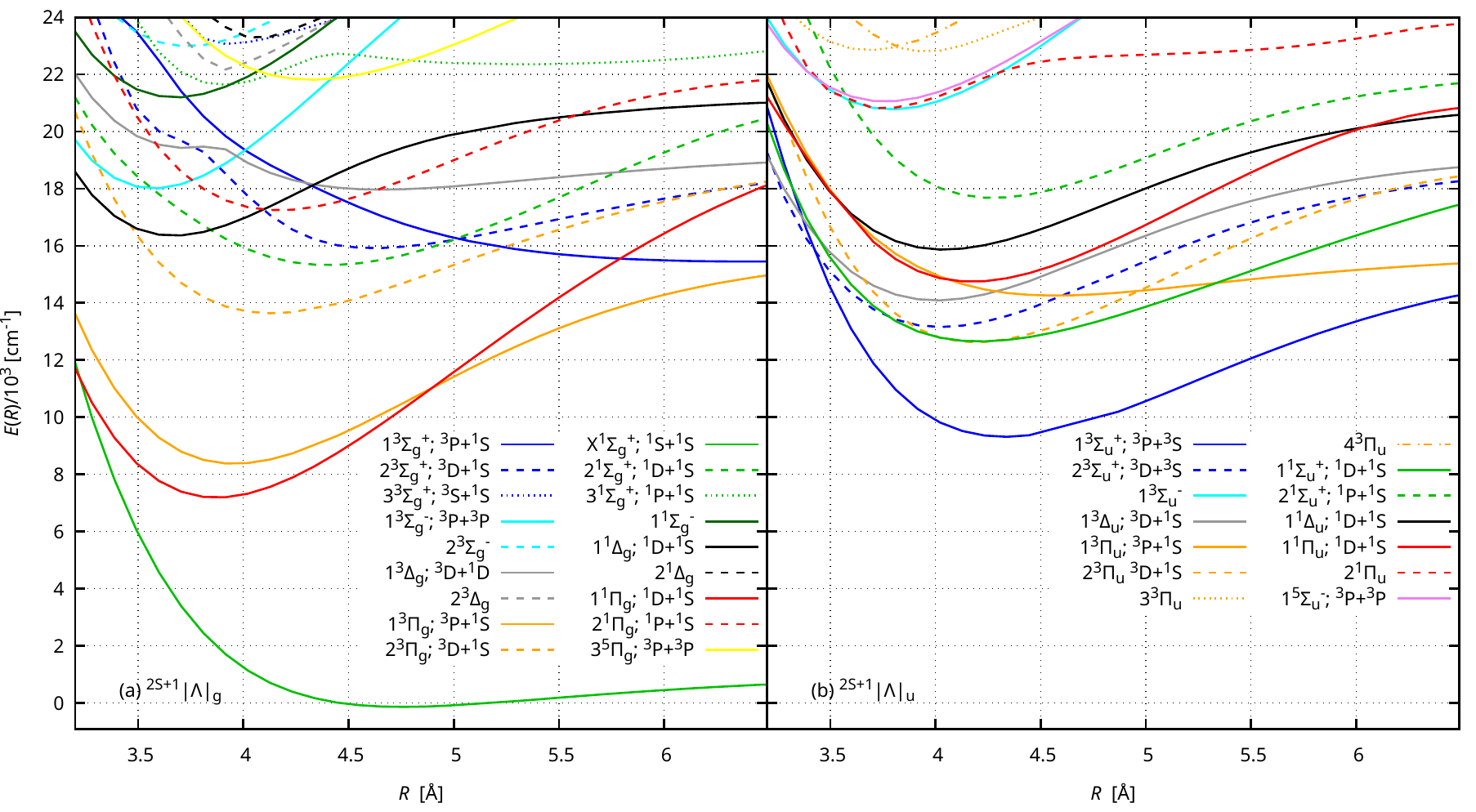}
\caption{Potential energy curves for the ground and excited electronic states of Sr$_2$ obtained in the non-relativistic spin-free sMRCI+Q/5Z computations with the scalar-relativistic small-core pseudopotential. The states are labeled by symmetry and asymptote or only by symmetry in the case of states involving asymptotes for which numerical difficulties prevent obtaining whole PECs.}
\label{fig:PECall}
\end{figure*}

\subsection{Theoretical results}

\begin{figure}
\includegraphics[width=\columnwidth]{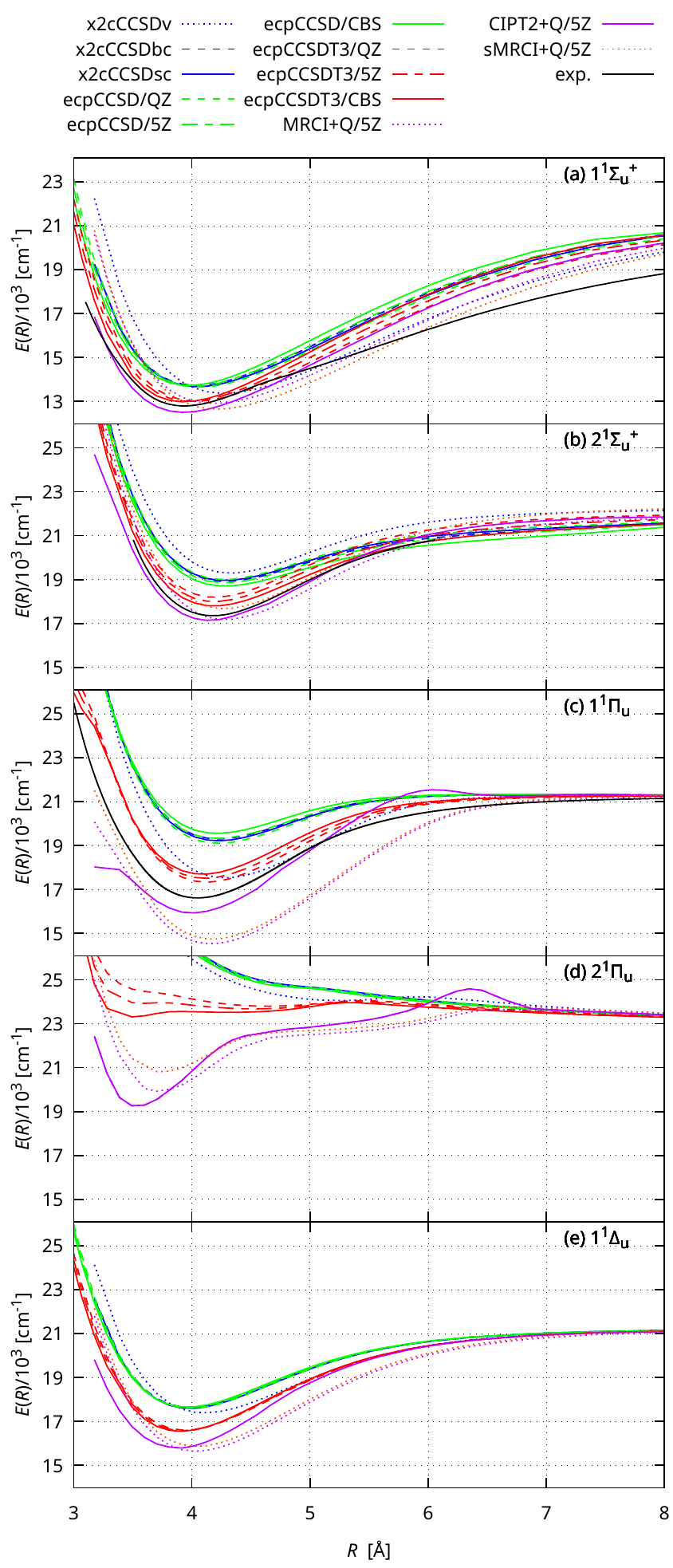}
\caption{Potential energy curves for the $1^1\Sigma_u^+$, $2^1\Sigma_u^+$, $1^1\Pi_u$, $2^1\Pi_u$, and $1^1\Delta_u$ electronic states of Sr$_2$ obtained with different computational methods. See the text for details. Thre potentials for $1^1\Sigma_u^+$, $2^1\Sigma_u^+$, and $1^1\Pi_u$ are compared with the experimental curves from the present work and Ref.~\cite{Stein2011}.}
\label{fig:from1P}
\end{figure}

Figure~\ref{fig:PECall} shows an overview of potential energy curves (PECs) computed with the sMRCI+Q/5Z approach, which is often the method of choice to study the excited electronic states of diatomic molecules. However, in the case of Sr$_2$, this approach has some shortcomings, as we shall discuss later. The main conclusion from the overview of excited states is that the $2^1\Sigma_u^+$ state is fairly well separated from other states, and, thus, a relatively small number of perturbations from other states should be expected.

The strontium dimer contains 76 electrons. Therefore, a precise quantum-mechanical description is challenging. Additionally, the large charge of its nuclei limits the applicability of non-relativistic quantum mechanics. Thus, an accurate description of Sr$_2$ has to include: \textit{i}) extensive orbital basis set, \textit{ii}) valence-electron correlation, \textit{iii}) core-electron correlation, \textit{iv}) scalar relativistic contribution, \textit{v}) spin-related relativistic effects, like fine and hyperfine couplings, and \textit{vi}) leading quantum electrodynamic corrections. However, it is currently only feasible to simultaneously account for some of these effects for many-electron molecules. Here, we neglect spin-related and quantum electrodynamic contributions (\textit{v}-\textit{vi}) and explore the sensitivity of the potential energy curves to the remaining contributions (\textit{i}-\textit{iv}), which are usually the most crucial for reaching quantitative description of any molecule. The spin-orbit coupling, the largest neglected contribution, can be perturbatively added in the next steps~\cite{TomzaPCCP11,SkomorowskiJCP12}. 

Figure~\ref{fig:from1P} presents the PECs for the $1^1\Sigma_u^+$, $2^1\Sigma_u^+$, $1^1\Pi_u$, $2^1\Pi_u$, and $1^1\Delta_u$ electronic states obtained at several different levels of theory. Corresponding spectroscopic parameters are collected in Tab.~\ref{tab:SpecParam}. We selected these singlet ungerade states for detailed computational tests because they are the most relevant for parallel spectroscopic measurements.

The first observation is that the PECs exhibit relatively low sensitivity to the orbital basis set size, meaning that calculations in the quadruple- and quintuple-zeta basis sets are already close enough to the complete basis set limit. This can be demonstrated by comparing the results in the quintuple-zeta basis set with the estimated CBS (ecpCCSDT3/5Z vs.~ecpCCSDT3/CBS). The difference in the dissociation energy is of the order of 30~cm$^{-1}$ for the $1^1\Sigma_u^+$ and $1^1\Delta_u$ states and of the order of 200~cm$^{-1}$ for the $1^1\Pi_u$ and $2^1\Sigma_u^+$ states.

Next, we observe that correlating only valence electrons is insufficient. We systematically investigate the effect of the core-electron correlation by changing the size of the frozen core in the all-electron equation-of-motion coupled-cluster computations with the eXact-2-Component Hamiltonian (x2cCCSDv vs.~x2cCCSDbc vs.~x2cCCSDsc). The lack of the core correlation not only alters the depth of PEC by hundreds of cm$^{-1}$, but mainly elongates the equilibrium distance. Surprisingly, the core correlation also affects the asymptotic region of the $1^1\Sigma_u^+$ and $2^1\Sigma_u^+$ states. On the other hand, the correlation of the valence and $4s4p$ electrons is sufficient, and the correlation of the electrons occupying the lower orbitals (replaced by the small-core pseudopotential) is not necessary.

Finally, we address the relativistic effects. Our calculations include the scalar relativistic effects only. We do not observe substantial differences between the all-electron x2cCCSDsc and small-core-pseudopotential ecpCCSD results. Therefore, we conclude that using the small-core ECP28MDF pseudopotential to account for the scalar relativistic effects is justified and sufficient, which is in agreement with other studies~\cite{SkomorowskiJCP12,TomzaMP13}. Our calculations do not account for the spin-related part of the Dirac-Coulomb-Breit Hamiltonian. It is necessary to go beyond this approximation to correctly describe the crossing between states of different spin multiplicity coupled by spin-orbit coupling, which can be added to our curves perturbatively. For example, this coupling is important for the $1^1\Sigma_u^+$ state at distances larger than 4.5$\,$\AA. Kotochigova~\cite{Kotochigova2008} included the spin-orbit part in her computations directly, but her results for $1^1\Sigma_u^+$ significantly deviate from modern experimental results and our calculations due to her approximate treatment of the electron correlation by the configuration interaction valence bond self-consistent-field approach. In contrast, Skomorowski \textit{et al.}~\cite{SkomorowskiJCP12} included the electron correlation directly and spin-orbit interaction perturbatively for $1^1\Sigma_u^+$ with the coupled cluster method and small-core pseudopotential, and obtained a much better agreement with the experiment.

\begin{table*}
\centering
\caption{The origin of the state $T_e$ with respect to the minimum of the ground electronic state, dissociation energy $E_e$, equilibrium bond length $R_e$, harmonic frequency $\omega_e$, equilibrium rotational constant $B_e$, and distortion constant $D_e$ for selected states of Sr$_2$ as predicted by various quantum chemical computations. These values are compared with experimental results whenever possible. The values in parentheses are experimental uncertainties in units of last digits.}
\begin{tabular}{cccccccc}
\hline \hline
state & method & $T_e$ & $E_e$ & $R_e$ & $\omega_e$ & $B_e$ & $D_e$ \\
 & & [cm$^{-1}$] & [cm$^{-1}$] & [\AA] & [cm$^{-1}$] & [cm$^{-1}$] & [$10^{-9}$ cm$^{-1}$] \\
\hline
$1^1\Sigma_u^+$ & ecpCCSDT3/QZ & 13013 & 8219 & 4.035 & 74.94 & 0.02356 & 9.32\\
$1^1\Sigma_u^+$ & ecpCCSDT3/5Z & 13010 & 8221 & 3.982 & 78.17 & 0.02419 & 9.27\\
$1^1\Sigma_u^+$ & ecpCCSDT3/CBS & 12984 & 8247 & 3.928 & 81.54 & 0.02486 & 9.24\\
$1^1\Sigma_u^+$ & ecpCCSD/5Z & 13693 & 7538 & 4.017 & 76.41 & 0.02376 & 9.20\\
$1^1\Sigma_u^+$ & sMRCI+Q/5Z & 12649 & 8582 & 4.209 & 71.25 & 0.02164 & 7.99\\
$1^1\Sigma_u^+$ & MRCI+Q/5Z & 12910 & 8321 & 4.181 & 71.12 & 0.02194 & 8.36\\
$1^1\Sigma_u^+$ & CIPT2+Q/5Z & 12502 & 8729 & 3.93 & 77.82 & 0.02484 & 10.0\\
$1^1\Sigma_u^+$ & x2cCCSDv & 13373 & 7858 & 4.272 & 69.86 & 0.02102 & 7.61\\
$1^1\Sigma_u^+$ & x2cCCSDbc & 13731 & 7500 & 4.037 & 75.28 & 0.02354 & 9.20\\
$1^1\Sigma_u^+$ & x2cCCSDsc & 13673 & 7559 & 4.044 & 74.81 & 0.02345 & 9.22\\

$1^1\Sigma_u^+$ & theory~\cite{Boutassetta1996} & 12363 & &3.850 & 79 & 0.0259 \\
$1^1\Sigma_u^+$ & theory~\cite{Czuchaj2003} & & 5490 & 4.01 & 80.21 & \\
$1^1\Sigma_u^+$ & theory~\cite{Kotochigova2008} & 17269 & 5475 & 4.02 & 88 & \\
$1^1\Sigma_u^+$ & theory~\cite{SkomorowskiJCP12} & & 8433 & 3.99 & \\
$1^1\Sigma_u^+$ & exp.~\cite{Stein2011} & 12796(2)& & 3.95(1) & 80.71(3) & 0.024794(2) \\
\hline
$1^1\Delta_u$ & ecpCCSDT3/QZ & 16605 & 4627 & 3.939 & 85.34 & 0.02472 & 8.30\\
$1^1\Delta_u$ & ecpCCSDT3/5Z & 16580 & 4652 & 3.921 & 85.74 & 0.02494 & 8.44\\
$1^1\Delta_u$ & ecpCCSDT3/CBS & 16550 & 4681 & 3.903 & 86.34 & 0.02518 & 8.56\\
$1^1\Delta_u$ & ecpCCSD/5Z & 17626 & 3605 & 3.974 & 80.18 & 0.02429 & 8.92\\
$1^1\Delta_u$ & sMRCI+Q/5Z & 15864 & 5367 & 4.041 & 85.22 & 0.02349 & 7.14\\
$1^1\Delta_u$ & MRCI+Q/5Z & 15646 & 5585 & 4.031 & 84.99 & 0.02361 & 7.28\\
$1^1\Delta_u$ & CIPT2+Q/5Z & 15786 & 5446 & 3.902 & 113.1 & 0.02519 & 5.00\\
$1^1\Delta_u$ & x2cCCSDv & 17393 & 3839 & 4.106 & 78.01 & 0.02275 & 7.74\\
$1^1\Delta_u$ & x2cCCSDbc & 17635 & 3596 & 3.982 & 79.82 & 0.02418 & 8.88\\
$1^1\Delta_u$ & x2cCCSDsc & 17595 & 3636 & 3.98 & 80.09 & 0.02421 & 8.84\\
$1^1\Delta_u$ & theory~\cite{Boutassetta1996} & 16158 & & 3.868 & 82 & 0.0257 \\
\hline
$1^1\Pi_u$ & ecpCCSDT3/QZ & 17331 & 3901 & 4.123 & 84.6 & 0.02256 & 6.42\\
$1^1\Pi_u$ & ecpCCSDT3/5Z & 17510 & 3722 & 4.107 & 84.67 & 0.02274 & 6.56\\
$1^1\Pi_u$ & ecpCCSDT3/CBS & 17695 & 3536 & 4.089 & 84.79 & 0.02293 & 6.71\\
$1^1\Pi_u$ & ecpCCSD/5Z & 19324 & 1907 & 4.217 & 75.01 & 0.02157 & 7.13\\
$1^1\Pi_u$ & sMRCI+Q/5Z & 14741 & 6490 & 4.173 & 85.77 & 0.02203 & 5.81\\
$1^1\Pi_u$ & MRCI+Q/5Z & 14530 & 6701 & 4.159 & 85.96 & 0.02217 & 5.90\\
$1^1\Pi_u$ & CIPT2+Q/5Z & 15933 & 5298 & 3.992 & 89.6 & 0.02407 & 6.94\\
$1^1\Pi_u$ & x2cCCSDv & 17530 & 3701 & 4.278 & 80.66 & 0.02096 & 5.66\\
$1^1\Pi_u$ & x2cCCSDbc & 19299 & 1932 & 4.226 & 75.57 & 0.02148 & 6.94\\
$1^1\Pi_u$ & x2cCCSDsc & 19219 & 2012 & 4.223 & 75.88 & 0.0215 & 6.91\\

$1^1\Pi_u$ & theory~\cite{Boutassetta1996} & 16243 & & 3.952 & 96 & 0.0246 & \\
$1^1\Pi_u$ & theory~\cite{Kotochigova2008} & 18658 & 4081 & 3.93 & 72 \\
$1^1\Pi_u$ & exp.~\cite{Stein2011} & 16617.86(2) & & 4.0473(2) & 86.300(3) & 0.023415(2) & 6.943 \\
\hline
$2^1\Sigma_u^+$ & ecpCCSDT3/QZ & 18193 & 4587 & 4.192 & 82.78 & 0.02183 & 6.07\\
$2^1\Sigma_u^+$ & ecpCCSDT3/5Z & 18000 & 4780 & 4.19 & 82.17 & 0.02185 & 6.18\\
$2^1\Sigma_u^+$ & ecpCCSDT3/CBS & 17797 & 4983 & 4.187 & 81.52 & 0.02187 & 6.30\\
$2^1\Sigma_u^+$ & ecpCCSD/5Z & 18858 & 3922 & 4.276 & 71.96 & 0.02097 & 7.13\\
$2^1\Sigma_u^+$ & sMRCI+Q/5Z & 17674 & 5106 & 4.273 & 82.66 & 0.021 & 5.42\\
$2^1\Sigma_u^+$ & MRCI+Q/5Z & 17182 & 5599 & 4.277 & 84.23 & 0.02096 & 5.19\\
$2^1\Sigma_u^+$ & CIPT2+Q/5Z & 17145 & 5635 & 4.145 & 88.54 & 0.02233 & 5.68\\
$2^1\Sigma_u^+$ & x2cCCSDv & 19297 & 3484 & 4.324 & 74.9 & 0.02051 & 6.16\\
$2^1\Sigma_u^+$ & x2cCCSDbc & 18914 & 3866 & 4.283 & 72.26 & 0.02091 & 7.01\\
$2^1\Sigma_u^+$ & x2cCCSDsc & 18959 & 3821 & 4.276 & 72.81 & 0.02097 & 6.96\\

$2^1\Sigma_u^+$ & present exp. & 17358.70(10) & & 4.176(1) & 84.217(3) & 0.021990(14) & 7.12(6)\\
$2^1\Sigma_u^+$ & exp.~\cite{Stein2011}& 17358.75(1) & & 4.1783(1) & 84.215(1) & 0.021969(1) & 5.98 \\
\hline
$2^1\Pi_u$ & CCSDT-3/CBS & 23846 & -1066 & 5.715 & 7.442 & 0.01174 & 117\\
$2^1\Pi_u$ & sMRCI+Q/5Z & 20805 & 1975 & 3.739 & 104 & 0.02743 & 7.64\\
$2^1\Pi_u$ & MRCI+Q/5Z & 19921 & 2859 & 3.677 & 137.1 & 0.02836 & 4.86\\
$2^1\Pi_u$ & CIPT2+Q/5Z & 19241 & 3539 & 3.533 & 142.9 & 0.03074 & 5.69\\
$2^1\Pi_u$ & present exp. & >19100 &  &  &  \\
\hline \hline
\end{tabular}
\label{tab:SpecParam}
\end{table*}

\begin{table*}[t]
\centering
\small
\caption{Excitation energies (in cm$^{-1}$) of singlet electronic states of the Sr atom obtained with  different quantum chemistry methods.}
\begin{tabular}{c|cccccc}
\hline \hline
method       & $^1D$     & $^1P$     & $^1S$     & $^1D$     & $^1P$     & $^1D$ \\ 
             & 5s4d      & 5s5p      & 5s6s      & 4d5p      & 5s6p      & 5s5d \\ 
\hline
exp.~\cite{NIST}         & 20149.685 & 21698.452 & 30591.825 & 33826.899 & 34098.404 & 34727.447\\
x2cCCSDv     & 20144     & 20817     & 29076     & 33569     & 32494     & 33623 \\
x2cCCSDbc    & 20659     & 22612     & 30982     & ---       & 34811     & 35355 \\
x2cCCSDsc    & 20855     & 22652     & 30993     & ---       & 34837     & 35405 \\
ecpCCSDT3/5Z & 20489     & 21845     & 30517     & 35776     & 34150     & 34820 \\
MRCI+Q/5Z    & 20162     & 20849     & 29095     & 33609     & ---   & --- \\
\hline \hline
\end{tabular}
\label{tab:AtomicExc}
\end{table*}

Overall, the primary factor determining the accuracy of the calculations for Sr$_2$ is the inclusion of high excitations in the description of the valence-electron correlation. For the analyzed electronic states, the difference between the ecpCCSDT3 and ecpCCSD results, that is, the inclusion of triple excitations, is more significant than the effect of the core-electron correlation. Among the methods used, the ecpCCSDT3 and CIPT2+Q approaches include the core-electron correlation and a substantial part of excitations higher than doubles. The CIPT2+Q method is a multireference approach that includes all possible excitations within the active space. Thus, CIPT2+Q accounts well for the static correlation but gives only an approximate description of the core-electron dynamic correlation. This approximation was necessary since an alternative approach, based on the multireference configuration interaction method with a large active space that correlates core electrons, goes beyond the technical capabilities of modern quantum-chemical programs. On the other hand, the single-reference ecpCCSDT3 approach accounts for the dynamic correlation of core and valence electrons but only includes a fraction of triple and higher excitations.

The importance of higher excitations and multireference nature can be seen by analyzing the wavefunctions. We inspected the squares of reference coefficients obtained with the sMRCI+Q method at $R=4.13\,$\AA, close to respective equilibrium distances. We found that the electronic wavefunction of the $1^1\Sigma_u^+$ state consists mostly of single-excited determinants (72$\,$\%), and the role of excitations higher than double is negligible (3$\,$\%). Therefore, it is not surprising that for this state, we observe the smallest difference between energies obtained with the ecpCCSDT3 and CIPT2+Q approaches, which also agree well with another single-reference calculation reported by Skomorowski \textit{et al.}~\cite{SkomorowskiJCP12}. The slightly smaller role of single-excited determinants is visible for the $2^1\Sigma_u^+$ and $1^1\Delta_u$ states (about 66\%), where the difference between PECs calculated with the ecpCCSDT3 and CIPT2+Q methods is larger. Still, the role of triple excitations for these states is below 5$\,$\%, and higher excitations are an order of magnitude less important. For the $1^1\Pi_u$ state, we observe similar contributions from single- and double-excited determinants (49$\,$\% and 39$\,$\%), that explains the significant difference between the ecpCCSD and ecpCCSDT3 results. The role of triple and quadruple excitations for this state is relatively low, accounting for about 5$\,$\% and 0.65$\,$\%, respectively. Therefore, we can conclude that the application of the full configuration interaction method may be unnecessary to obtain accurate results for the $1^1\Sigma_u^+$, $2^1\Sigma_u^+$, $1^1\Delta_u$, and $1^1\Pi_u$ states. We can also assume that the ecpCCSDT3 and CIPT+Q approaches properly set the boundaries for the PEC shapes. Indeed, near the minima, the experimental PECs for the $1^1\Sigma_u^+$, $2^1\Sigma_u^+$, and $1^1\Pi_u$ states lie between the curves obtained with the ecpCCSDT3 and CIPT2+Q methods.

The $2^1\Pi_u$ state deserves special attention and particular comment because the variation of PECs obtained for this state with different methods is the largest, and this is the only state analyzed in which double excitations are dominant (75$\,$\%). Additionally, it exhibits the highest contribution from the quadrupole excitations (2\%). It corresponds to the $^1P$+$^1S$ asymptote, and for large internuclear distances, it is repulsive. We can observe its two avoided crossings with other states of the same symmetry. The one at a larger interatomic separation involves a state from the $^3P$+$^3P$ asymptote. The assignment of the second crossing is far more complex, as numerical difficulties prevent obtaining whole PECs for all states from the $^1D(4d5p)$+$^1S(5s^2)$, 
$^1P(5s5p)$+$^1S(5s^2)$,
$^1D(5s5d)$+$^1S(5s^2)$ and 
$^3D(5s4d)$+$^3P(5s5p)$ asymptotes. We suppose that doubly excited states, $^1D(4d5p)$+$^1S(5s^2)$ and $^3D(5s4d)$+$^3P(5s5p)$, play a crucial role here. The equation-of-motion coupled cluster method with single and double excitations does not describe them accurately, so the PEC predicted at that level is mostly repulsive. The inclusion of some higher excitations by the ecpCCSDT3 method allows for a poor description of $^1D(4d5p)$ state of Sr, where the excitation energy is overestimated by nearly 2000~cm$^{-1}$ (see Tab.~\ref{tab:AtomicExc}). The MRCI+Q/5Z method predicts the energy of $^1D(4d5p)$ in a reasonably good agreement with the experiment. On the other hand, MRCI+Q/5Z tends to overestimate the depth of the potential for other states of Sr$_2$. Additionally, our active space is too small to fully account for the $^1P(5s5p)$+$^1S(5s^2)$ and $^1D(5s5d)$+$^1S(5s^2)$ asymptotes. Overall, the expected position of the minimum in the $2^1\Pi_u$ potential is in the wide range between 19000$\,$cm$^{-1}$ given by the CIPT2+Q/5Z method and 24000$\,$cm$^{-1}$ from the ecpCCSDT3/CBS computation above the minimum of the ground electronic state (see Tab. \ref{tab:SpecParam}). This range covers the value $T_e>19100$ cm$^{-1}$ estimated based on our present experimental observations (\textit{vide infra}). Our computations do not confirm the existence of the avoided crossing between $1^1\Pi_u$ and $2^1\Pi_u$ predicted by Boutassetta \textit{et al.}.~\cite{Boutassetta1996} We indeed observe that $1^1\Pi_u$ and $2^1\Pi_u$ approach each other in the repulsive part of the PECs, but a possible crossing may occur only in the region experimentally insignificant. We suppose that a small basis set, with an insufficient number of high angular momentum components, could have significantly decreased the precision of Boutassetta \textit{et al.} results for highly excited states. However, their approach accounts well for static correlation and thus reproduces the general shape of the PECs. Czuchaj \textit{et al.}~\cite{Czuchaj2003} reported PEC, which differs from our and Boutassetta \textit{et al.} results. However, their active space in multireference computations was smaller than ours and did not allow for an accurate description of static correlation. We believe that the accurate description of $2^1\Pi_u$ state is a major computational challenge. Most likely, the only way to obtain its reliable and accurate description is to use the full configuration interaction with a large basis set and proper account for the core and core-valence correlation method, which is out of the scope of this work. Therefore, for now, we must assume that the exact shape of the curve is unknown and falls somewhere between the curves predicted by the ecpCCSDT3/CBS and CIPT2+Q/5Z approaches. 

We provide the potential energy curves for the $1^1\Sigma_u^+$, $2^1\Sigma_u^+$, $1^1\Pi_u$, $2^1\Pi_u$, and $1^1\Delta_u$ states obtained with the most-accurate ecpCCSDT3/CBS and CIPT2+Q/5Z methods, and for the states presented in Fig.~\ref{fig:PECall} obtained with the sMRCI+Q/5Z method in the supplementary data accompanying this paper~\cite{supplC}.

\section{Experiment}
\label{sec:experiment}
\subsection{Experimental setup}

The Sr$_{2}$ molecules were produced in a three section heat-pipe oven~\cite{Ciamei2018} of 1~m length filled in the central part with 15~g of strontium. The central part with a length of 20~cm was heated to 1020$^{\circ}$C, while external parts were maintained at 720$^{\circ}$C. In the case of strontium a proper circulation of the metal inside the heat-pipe is a challenge. To solve this problem, 1.5~g of metallic magnesium was added to its central section (strontium and magnesium form an alloy with substantially lower melting point than its constituents~\cite{MgSr_NayebHashemi1986}). A steel mesh was placed separately in each section and the heat-pipe was filled with 15~Torr of argon buffer gas. 

A polarisation labelling spectroscopy (PLS) technique was employed to obtain spectra of Sr$_{2}$ molecules. The PLS is a pump-probe experimental technique, which takes advantage of an optical anisotropy created in a chosen group of molecules in the sample to limit the number of observed spectral lines~\cite{Szczepkowski2013}. In the present experiment a NarrowScan dye laser of a spectral linewidth of 0.07~cm$^{-1}$ pumped with a XeCl excimer laser (Light Machinery) was used as a pump laser and its wavelength was scanned between 18400~cm$^{-1}$ and 20300~cm$^{-1}$, covering transitions from the ground state of Sr$_2$ to the upper part of the $2^1\Sigma^{+}_{u}$ state. As a probe laser a ring dye laser (Coherent 899, pumped with Sprout laser) was employed, and its wavelength was controlled with HighFinesse WS-7 wavemeter. The laser was working on Rhodamine 6G, what enabled tuning its light within a spectral range 16800~cm$^{-1}$ -- 17600~cm$^{-1}$. The probe laser wavelength was fixed on selected transitions from the ground X$^1\Sigma^{+}_{g}$ state of Sr$_{2}$ molecule to low levels of the $2^1\Sigma^{+}_{u}$ state known from supplementary materials of the publication describing the bottom part of this state~\cite{Stein2011}. 

Reference signals for wavenumber calibration of the molecular spectra were needed, therefore two auxiliary signals were recorded, namely transmission fringes from a Fabry-P\'{e}rot interferometer with $\mathrm{FSR} = 1\,\mathrm{cm^{-1}}$, and  optogalvanic spectrum from argon and neon hollow-cathode lamps. This ensured that the uncertainty of wavenumbers determined in this way is below $\pm 0.1\,\mathrm{cm^{-1}}$. 

\subsection{Analysis of the spectra}

\begin{figure}
\includegraphics[width=0.99\linewidth]{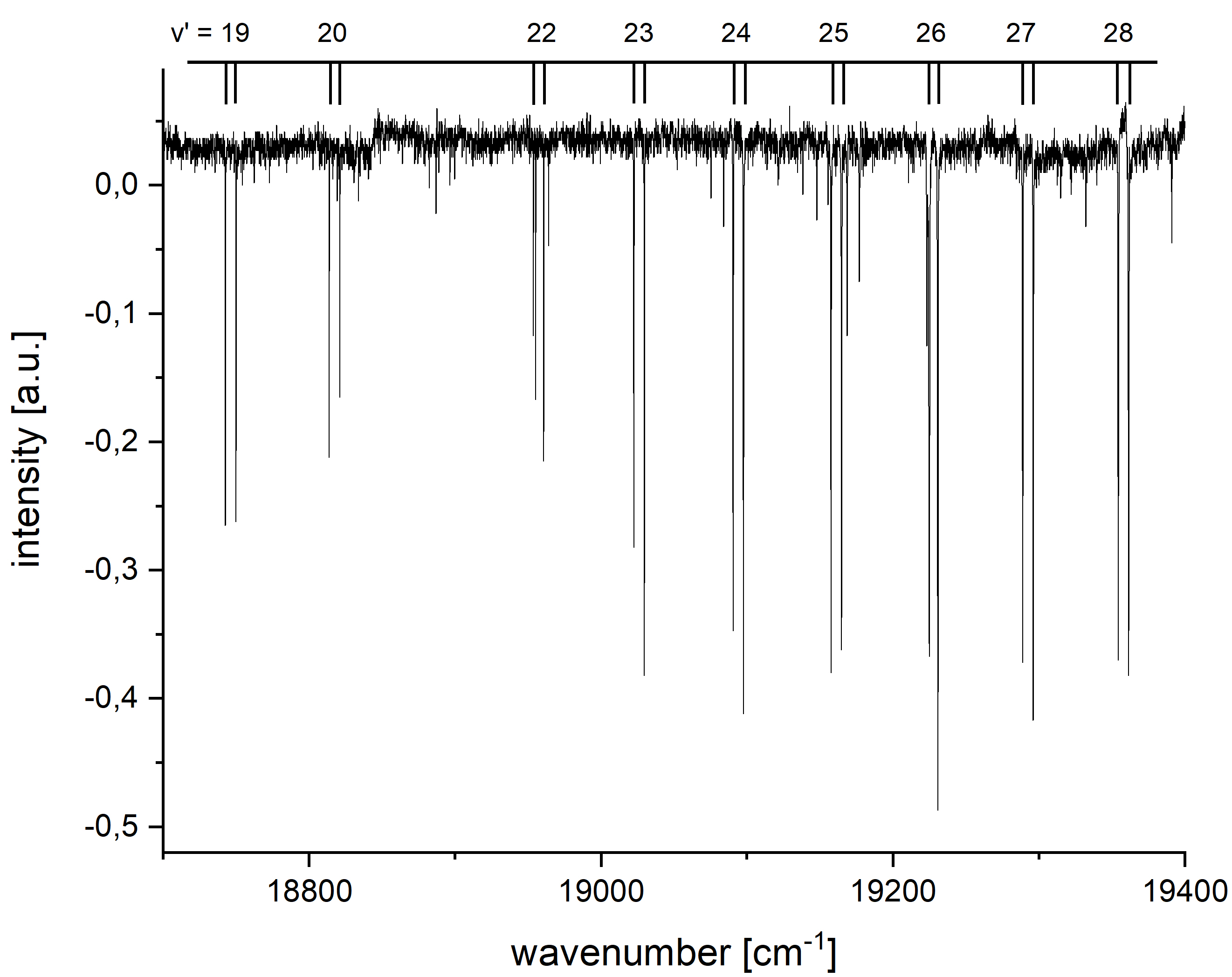}
\caption{A portion of the experimental spectrum of Sr$_2$ coresponding to transitions from rovibrational level $v''=4$, $J''=90$ in the ground X$^1\Sigma^{+}_{g}$ state  to consecutive rovibrational levels ($v'=19-28$) of the excited $2^1\Sigma^{+}_{u}$ state. }
\label{fig:spectrum}
\end{figure}

As the bottom part of the $2^1\Sigma^{+}_{u}$ state has been characterised in the Fourier-transform spectroscopy experiment by Stein \textit{et al.}~\cite{Stein2011}, we have concentrated on higher vibrational levels of this state. A typical example of the recorded spectrum of Sr$_2$ is presented in Fig.~\ref{fig:spectrum}. Our experiment provided information about rovibrational levels with quantum numbers ranging from $v'=13$ to 52 and $J'$ from 43 to 149, solely in the most abundant $^{88}$Sr$_2$ isotopologue. We supplemented the database with levels of $^{88}$Sr$_2$ measured in~\cite{Stein2011}, however to avoid too strong influence of these levels on subsequent fits of molecular constants we limited the borrowed levels to $J'<150$ and assigned them with the same accuracy of 0.1~cm$^{-1}$ as our own data. This resulted in 714 levels taken from~\cite{Stein2011} and 1760 levels from our own measurements. The term values of all levels were calculated by adding the measured transition energies to the energies of the initial X$^1\Sigma^{+}_{g}$ ($v''$, $J''$) levels obtained from the highly accurate molecular constants reported in Ref.~\cite{Stein2008}.

\begin{figure}[tb]
	\includegraphics[width=0.99\linewidth]{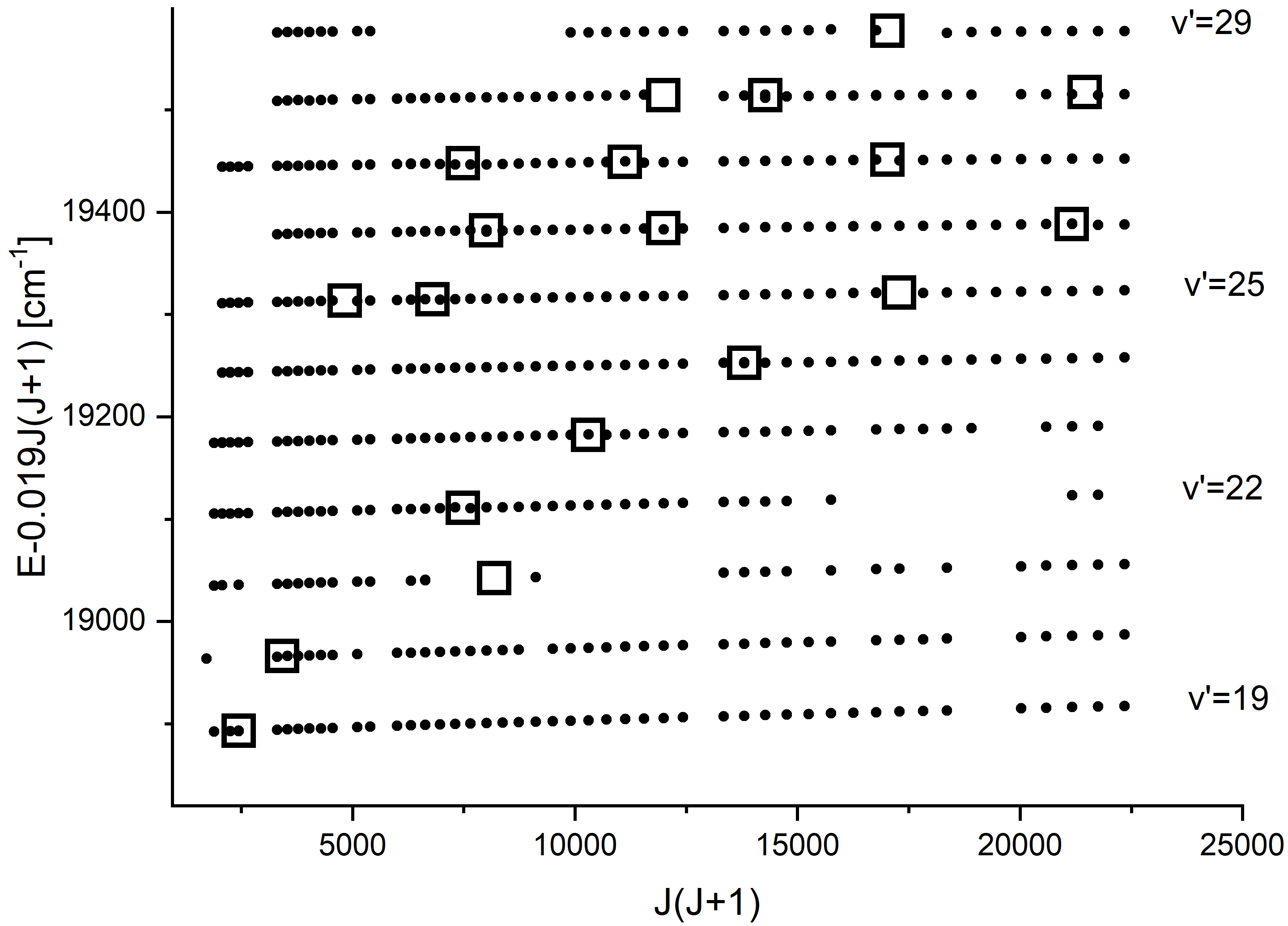}
	\caption{Reduced term values $E_\mathrm{red}=E-0.19 \times J(J+1)$~cm$^{-1}$ of part of the observed rovibrational levels in the $2^1\Sigma^{+}_{u}$ state (dots) plotted against $J(J+1)$. Open squares indicate approximate positions of centres of perturbations.}
	\label{fig:pert}
\end{figure}

\begin{figure}[tb]
	\includegraphics[width=0.99\linewidth]{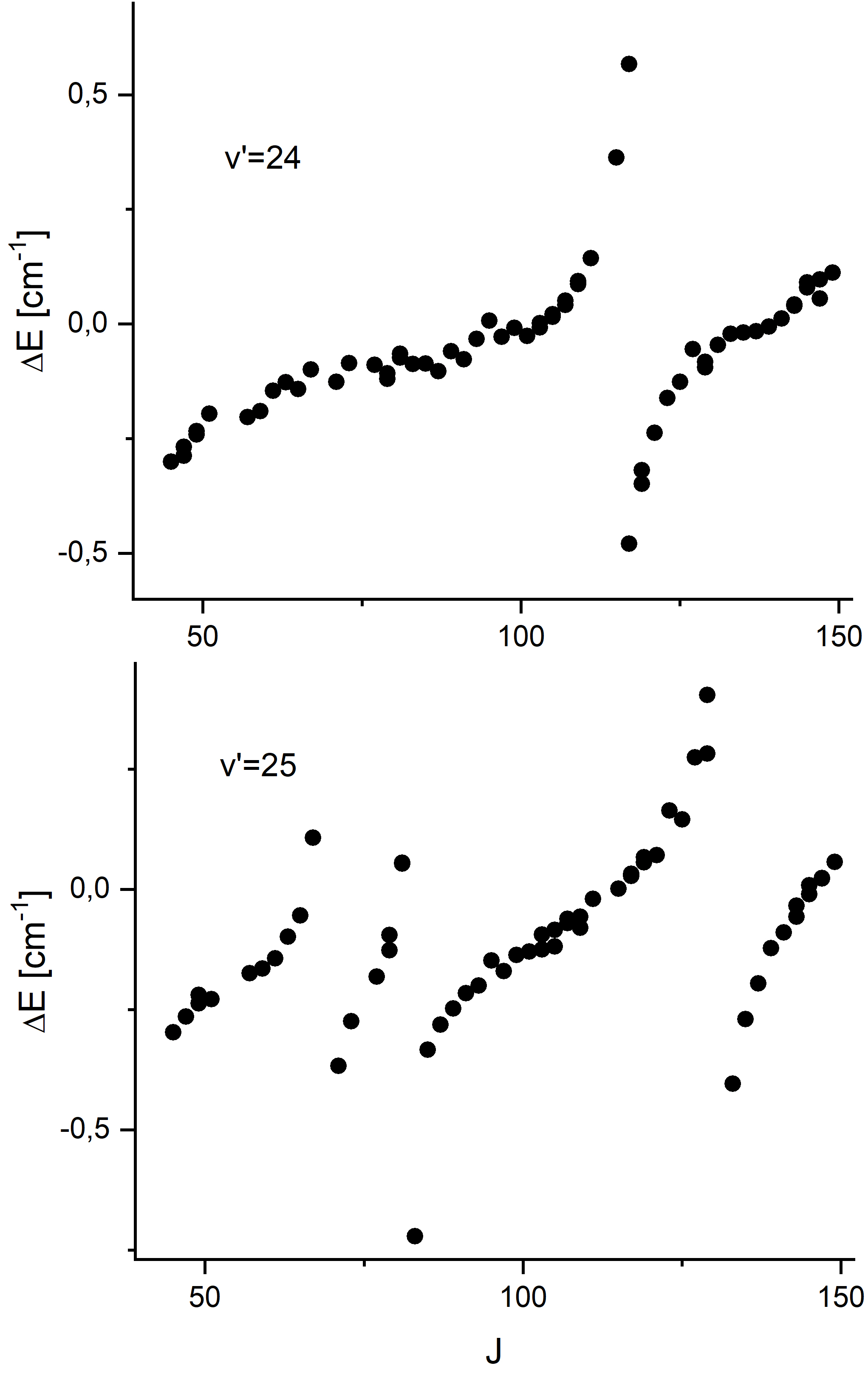}
	\caption{Observed shifts of the rotational energy levels in the $2^1\Sigma^{+}_{u}$ state from their predicted positions for vibrational levels $v'=24$ and 25.}
	\label{fig:lsh}
\end{figure}

Originally we tried to fit the term energies to the standard Dunham expansion
\begin{equation}
T(v,J)=T_e+\sum_{m,n}Y_{mn}(v+1/2)^{m}J(J+1)]^{n}\,,
 \label{Dunh}
\end{equation}
but the rms error of the fit amounted to 0.5~cm$^{-1}$, i.e. five times more than our experimental accuracy. This result suggested strong perturbations in the $2^1\Sigma^{+}_{u}$ state, particularly that the misbehaving levels, all of them corresponding to $v'> \approx19$, were centred around isolated ($v'$, $J'$) values and the deviations fell into patterns characteristic for perturbations. Figure~\ref{fig:pert} displays term values of several levels of the $2^1\Sigma^{+}_{u}$ state plotted against $J(J+1)$ and open squares localise regions of the observed perturbations. Figure~\ref{fig:lsh} visualises typical pattern of deviations between the observed and predicted line positions versus rotational quantum number $J$.

In the preliminary analysis presented here which aims primarily to test accuracy of theoretical predictions based on various computational methods, we decided to remove the apparently perturbed levels from the database and to fit Dunham coefficients to the remaining levels. When fitting energies of somewhat arbitrarily chosen 1636 levels (out of the total 2474) we obtained rms error 0.08~cm$^{-1}$, this time consistent with the experimental accuracy. The Dunham coefficients have been rounded to minimise the number of digits by a procedure described by Le Roy~\cite{Roy1998}. They are listed in Tab.~\ref{tab:Dunhams} together with the equilibrium bond length $R_e$ calculated from the rotational constant and reduced mass of strontium nuclei. A rotationless potential energy curve for the $2^1\Sigma^{+}_{u}$ state was constructed by a standard Rydberg-Klein-Rees (RKR) method. The vibrational term energies $G_v$ and turning points $R_-$ and $R_+$ are given in Tab.~\ref{tab:RKR} and the potential curve is displayed in Fig.~\ref{fig:from1P} along with the theoretical predictions. Our work extends the range of experimentally determined potential to 3.5 \AA\ $< R <$ 6.1 \AA\ and more than doubles the range of covered energies.

\begin{table}[]
\centering
\caption{The Dunham coefficients that describe the $2^1\Sigma^{+}_{u}$ state of $^{88}$Sr$_2$ in the range $0 \leq v' \leq 52$, $J' \leq 149$. The
numbers in parentheses give uncertainties in the last quoted digits (one standard deviation).}
\begin{tabular}{cc}
\hline \hline
constant       & value [cm$^{-1}$]      \\ 
\hline
$T_e$   &   17358.70(10) \\
$Y_{10}$   &   84.2169(27) \\
$Y_{20}$   &   -0.26729(21) \\
$Y_{30}$   &   -0.10850(6) $\times 10^{-2}$ \\
$Y_{40}$   &   -0.10760(6) $\times 10^{-4}$ \\
$Y_{01}$   &   0.021990(14) \\
$Y_{11}$   &   -0.6808(18) $\times 10^{-4}$ \\
$Y_{21}$   &   -0.4860(10) $\times 10^{-6}$ \\
$Y_{31}$   &   -0.1030(13) $\times 10^{-7}$ \\  
$Y_{02}$   &   -0.712(6) $\times 10^{-8}$ \\
\hline
$R_e$ [\AA] &  4.176(1) \\
\hline \hline
\end{tabular}

\label{tab:Dunhams}
\end{table}

\begin{table*}[]
\centering
\caption{The rotationless RKR potential energy curve for the $2^1\Sigma^{+}_{u}$ state of Sr$_2$. The first line refers to the bottom of the potential curve, the corresponding $R$ value is the equilibrium distance. The full list of turning points of the potential is given in supplementary materials accompanying this work~\cite{supplC}.}
\begin{tabular}{ccccccccc}
\hline \hline
$v$  & $G_v$ [cm$^{-1}$] & $R_-$ [\AA] & $R_+$ [\AA] & &  $v$  & $G_v$ [cm$^{-1}$] & $R_-$ [\AA] & $R_+$ [\AA]    \\ 
\hline
  &  0  & 4.176 & \\
0	& 42.024	& 4.084	& 4.275  & & 26	& 2018.540	& 3.630	& 5.153  \\
1	& 125.703	& 4.020	& 4.352  & & 28	& 2150.854	& 3.616	& 5.210  \\
2	& 208.838	& 3.978	& 4.408  & & 30	& 2279.868	& 3.603	& 5.268  \\
4	& 373.445	& 3.917	& 4.496  & & 32	& 2405.468	& 3.590	& 5.327  \\
6	& 535.786	& 3.870	& 4.571  & & 34	& 2527.538	& 3.579	& 5.388  \\
8	& 695.797	& 3.832 & 4.638  & & 36	& 2645.955	& 3.568	& 5.451  \\
10	& 853.411	& 3.799	& 4.701  & & 38	& 2760.596	& 3.558	& 5.516  \\
12	& 1008.555	& 3.770	& 4.760  & & 40	& 2871.329	& 3.549	& 5.583  \\
14	& 1161.155	& 3.744	& 4.818  & & 42	& 2978.022	& 3.540	& 5.653  \\
16	& 1311.129	& 3.721	& 4.875  & & 44	& 3080.537	& 3.532	& 5.726  \\
18	& 1458.395	& 3.700	& 4.930  & & 46	& 3178.731	& 3.524	& 5.803  \\
20	& 1602.863	& 3.680	& 4.986  & & 48 & 3272.459	& 3.517	& 5.884  \\
22	& 1744.442	& 3.662	& 5.041  & & 50	& 3361.570	& 3.510	& 5.970  \\
24	& 1883.034	& 3.646	& 5.097  & & 52	& 3445.910	& 3.504	& 6.061  \\

\hline \hline
\end{tabular}

\label{tab:RKR}
\end{table*}

It must be noted that in the range $v' = 0$ -- $\approx18$ the $2^1\Sigma^{+}_{u}$ state is free of (strong) perturbations which become visible only from $v'\approx19$. The most likely perturber is the $1^1\Pi_u$ state, as the outer limb of its potential curve gradually approaches potential of the $2^1\Sigma^{+}_{u}$ state. However, at $v'\approx25$ apparently an additional perturbing state emerges, since perturbations become more frequent ($v'=25$ is perturbed at least around four $J'$ values, see Fig.~\ref{fig:lsh}). From analysis of the theoretical potential energy curves it follows that the new perturber must be the $2^1\Pi_u$ state, the only other singlet ungerade state which is expected to be present in the vicinity. Therefore our observation indicates that the bottom of its potential cannot be located higher than approximately 19100~cm$^{-1}$ above the minimum of the ground state potential, what can serve as another test for validity of theoretical predictions. 

\section{Summary and conclusion}
\label{sec:conclusion}

In this study, we investigated the excited electronic states of the strontium dimer. We theoretically obtained the complete molecular electronic spectrum up to the excitation energy of around 25000$\,$cm$^{-1}$ using the multireference configuration interaction method. Next, we studied in detail potential energy curves for the $1^1\Sigma^{+}_{u}$, $2^1\Sigma^{+}_{u}$, $1^1\Pi_{u}$, $2^1\Pi_{u}$, and  $1^1\Delta_{u}$  states using several advanced electronic structure method. We evaluated the importance of the orbital basis set size, core- and valence-electron correlation, and scalar relativistic effects. Theoretical results had been motivated by our ongoing spectroscopic studies. We presented new experimental measurements of the excited $2^1\Sigma^{+}_{u}$ state using polarisation labelling spectroscopy, extending the range of observed vibrational levels to higher energies. We reported the corresponding Dunham coefficients and experimental potential energy curve. We observed perturbations in the recorded spectrum that give preliminary information on higher-excited electronic states. We compared the available experimental observations with the theoretical predictions to assess the accuracy and limitations of employed theoretical models. Our findings provide valuable insights into the complex electronic structure of Sr$_2$, paving the way for future, more accurate theoretical and experimental spectroscopic studies.

The challenging nature of excited electronic states of the Sr$_2$ dimer makes them a perfect testbed and playground for near-future developments of the electronic structure theory and computation. In the following work, we plan to present calculations at the valence full configuration interaction level with large basis sets. Such converged calculations in versions with both small-core and large-core pseudopotentials and approximately included core and core-valence correlation may resolve the nature of most problematic states such as $2^1\Pi_u$. 

A more rigorous deperturbation procedure is also needed to clarify the experimental observations. It would require a coupled channels treatment involving (at least) the $2^1\Sigma^{+}_{u}$, $1^1\Pi_u$ and $2^1\Pi_u$ interacting states and such an analysis is in future plans of our group. However, more precise theoretical predictions of the relevant potential energy curves are needed to serve as a starting point for deperturbation procedure. At the present stage we show the approximate results, accuracy of which is more than sufficient for comparison with theoretical models, and all the experimental term energies are listed in the supplementary data accompanying this paper~\cite{supplC}.

\section*{Acknowledgements}
Financial support from the National Science Centre Poland (grant no.~2021/43/B/ST4/03326) is gratefully acknowledged. The computational part of this research has been partially supported by the PL-Grid Infrastructure (grant no.~PLG/2021/015237).

\bibliography{Sr2}

\end{document}